\documentclass{PoS}

\usepackage{epsf}
\usepackage{psfig}

\title{The evolving starburst-AGN connection: Implications for SKA and its pathfinders}

\ShortTitle{The evolving starburst-AGN connection}

\author{\speaker{Ray P. Norris}\\
         CSIRO ATNF\\
         E-mail: \email{Ray.Norris@csiro.au}}

\author{ Enno Middelberg\\
        University of Bochum\\
        E-mail: \email{middelberg@astro.rub.de }}

\author{ Brian J. Boyle\\
        CSIRO ATNF\\
        E-mail: \email{Brian.Boyle@csiro.au }}

\abstract{ How well is the modern-day starburst-AGN connection mirrored in the early Universe? This is starting to be answered by deep wide radio surveys such as ATLAS, which are giving us a new view of  high redshift galaxies. For example, we find powerful radio-loud AGNs which look like star-forming spirals in the optical and infrared, a composite which is almost unknown in the modern Universe. We find radio-bright objects which are unexpectedly invisible in the infrared, and which may be very high redshift radio galaxies and quasars. And although the radio-far-infrared correlation for star-forming galaxies has now been extended down to microJy levels, we still cannot reliably distinguish between starburst and AGN. So what do we need to do to ensure that SKA and its pathfinders will be able to understand galaxy evolution in the early Universe?
}

\FullConference{From Planets to Dark Energy: the Modern Radio Universe\\
 		 October 1-5 2007\\
 		 The University of Manchester, UK}

\begin{document}


\pagestyle{myheadings}
\setcounter{equation}{0}
\setcounter{figure}{0}
\setcounter{footnote}{0}
\setcounter{section}{0}
\setcounter{table}{0}

\section{Introduction}
Deep radio surveys are becoming increasingly important for studying galaxy evolution, because they are unaffected by dust and can detect objects up to the highest redshifts. The Australia Telescope Large Area Survey (ATLAS) is surveying a seven square degree area with the aim of producing the widest deep (10-15 $\mu$Jy rms) radio survey ever attempted.

The power of deep radio surveys is often limited by inadequate data at other wavelengths. This is overcome in ATLAS by choosing fields observed by the Spitzer Wide-area Infrared Extragalactic Survey (SWIRE) program (Lonsdale et al. 2003), around the CDFS and ELAIS-S1 regions. As a result, the ATLAS radio data are accompanied by extensive radio, infrared, and optical data. 

The key science goals of ATLAS are: 
\begin {itemize}

\item To determine the relative contribution of starbursts and AGN to the overall energy density of the universe, and the relationship between AGN and star-forming activity. 

\item To test whether the radio-far-infrared correlation changes with redshift or with other galaxy properties, and measure the cosmic star formation history. 

\item To study large scale structure and clustering, using the radio sample which will be unbiased by dust. This will include searching for over-densities of high-z ULIRGs which mark the positions of proto-clusters in the early Universe. 

\item To trace the radio luminosity function to a high redshift, and accurately measure for the first time the differential 20 cm source count to a flux density limit of  about 30 $\mu$Jy. 

\end {itemize}

In addition to these astrophysical goals, surveys such as ATLAS, which are breaking into new areas of observational phase space, are also important trailblazers for the much more extensive studies that will be possible with SKA and its pathfinders. In this paper, we highlight the lessons learnt from this work that are relevant to the planning of next-generation radio-telescopes.

\section{Observations}

We are currently part way through the ATLAS observations, having covered 7 square degrees of the CDFS and ELAIS-S1 SWIRE fields to a sensitivity of about 30-40 $\mu$Jy, and a spatial resolution of about 10 arcsec. From these images we have identified and catalogued about 2000 radio sources, and have made the data products publicly available (Norris et al. 2006, Middelberg et al. 2008). When the survey is complete, we hope to reach an rms of 10-15 $\mu$Jy over this field. A typical part of the field is shown in Fig. 1.

Nearly all the radio sources have been detected by Spitzer, and many have optical photometry, giving up to 10 bands of photometry for SED fits and photometric redshifts. We also have a spectroscopic program, as well as observations at other wavelengths. 

\begin{figure}[]
\begin{center}
\epsfysize=50mm
\epsfbox[0 10 1000 400]{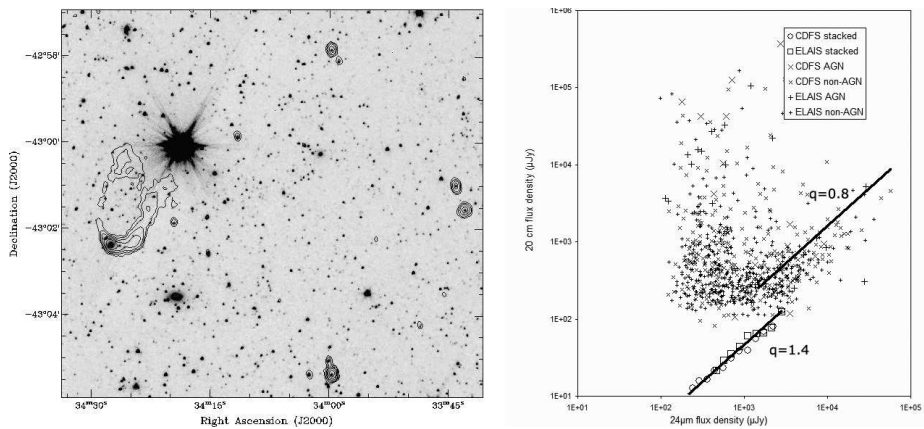}
\caption{(Left) A small part of the radio image, showing radio contours overlaid on the 3.6$\mu$m SWIRE image, showing a prominent Wide-Angle Tail galaxy, together with other AGN and star-forming galaxies. (Right) The flux densities of sources in the two ATLAS fields, together with stacked data (Boyle et al. 2007).}
\end{center}
\end{figure}

Although the rms noise in the ATLAS images is largely close to the theoretical limit, it is degraded in some places by imaging artefacts which are resistant to standard tools such as cleaning, selfcal, and peeling. Instead, they appear to be caused by the non-circularity of the primary beam. Such effects have probably not been encountered before because the ATLAS project is the first to survey such a large area to this depth. We plan to develop software and algorithms to correct for this, because it not only affects ATLAS, but will also be a major challenge for next-generation radio telescopes such as ASKAP (Deboer et al., 2007) and SKA (Carilli \& Rawlings, 2004). 

\underline{SKA Lesson 1:} {\em As we probe deeper in dynamic range and sensitivity, expect to find new imaging challenges and artefacts which may prove expensive to overcome.}

\section{The radio-FIR correlation}
Low-redshift non-AGN galaxies show a tight correlation between the far-infrared and radio emission (e.g. de Jong et al. 1985), which is reflected in a similar correlation between the 24 $\mu$m and 20 cm emission. Appleton et al. (2004) have shown this correlation extends out to z $\sim$ 1, and Beswick et al. (2007) have shown that the luminosities of all identified 24 $\mu$m sources still follow the correlation down to S$_{24{\mu}m} \sim 80 \mu$Jy. To explore the correlation at even fainter levels, Boyle et al. (2007) stacked ATLAS radio data to show that it extends down to microJy levels.

Fig. 1 shows all ATLAS sources with Spitzer 24 $\mu$m flux densities. They include both star-forming galaxies, which roughly follow the correlation found by Appleton et al., and AGN, which extend to the top of the plot.
Also shown are the stacked data from Boyle et al. Although the stacked data broadly follow the correlation, they follow a value of q=log(S $_{24\mu m}$/S$_{20cm}$)=0.8 which is significantly different from the q=1.4 found by Appleton et al. We have found (Norris et al., 2008) that this discrepancy is not an artefact of the processing, occurs in other radio surveys, and is attributable to sources with a low measured radio flux density.

\section{Distinguishing AGN from star-formation activity}

Attempts to understand the evolution of galaxies continue to be plagued by the problem of distinguishing between star-formation activity and AGN. While optical emission line ratios remain the standard discriminant for low-redshift galaxies, they are unreliable for dusty galaxies, which constitute a substantial fraction of galaxies in deep radio surveys.

We are therefore exploring the use of other indicators of AGN and star forming activity, such as optical/infrared spectral energy distributions (SEDs), spectroscopic line widths, radio spectral indices, radio variability, radio morphology, radio polarization, departure from the radio/far-infrared relation, and the presence of VLBI cores.
While none of these on its own is unambiguous, together they constitute a valuable toolkit for distinguishing between AGN and star-forming activity. However, it is becoming clear that the challenge is not to assign each galaxy to a star-forming category or an AGN category, but to ask what fraction of the luminosity of a galaxy is produced by the AGN. 

\section{Obscured AGN}

Radio observations are unaffected by dust, and so enable us to detect obscured AGN. For example, Fig. 2 shows the source S425, which is a classic triple radio source at a photometric redshift of 0.932, with a luminosity of $4 \times 10^{25}$ WHz$^{-1}$, placing it close to the FRI/FRII break.

However, the host galaxy has the SED of a spiral galaxy. The ATLAS catalogue includes several examples of galaxies whose SED is of a star-forming spiral galaxy, but which have the radio luminosity or morphology of an AGN. We suggest that these represent a class of AGN buried deeply inside a dusty star-forming galaxy, which appear to be increasingly common at high redshifts.

Such objects are rare in the local Universe, with only one nearby example: 0313-192 (Keel et al. 2006). An optical survey would classify this source as a star-forming galaxy, demonstrating the value of radio data in understanding the cosmic evolution of AGN and star formation.

\underline{SKA Lesson 2:} {\em Don't rely too much on the local universe to tell you what happens at high z.}

\begin{figure}[]
\begin{center}
\epsfysize=50mm
\epsfbox[0 10 1000 400]{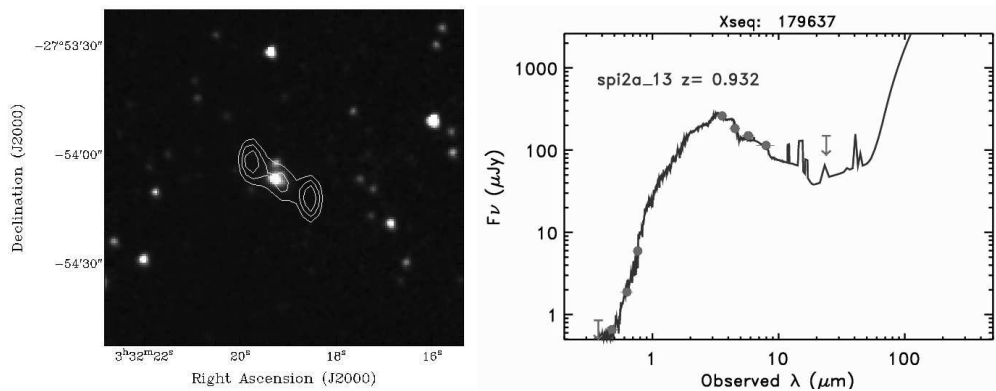}
\caption{The composite source S425. The left image shows radio contours superimposed on a 3.6 $\mu$m SWIRE image, and the right image (Mari Poletta, private communication) shows the SED. This source has the radio morphology of an AGN, but the SED of a star-forming galaxy.}
\end{center}
\end{figure}

\section{Infrared-Faint Radio Sources (IFRS)}

Unexpectedly, we have found 53 radio sources (IFRS: Infrared-Faint Radio Sources) as strong as 20 mJy at 20 cm, which are undetected by Spitzer. Norris et al (2006) have shown that they are undetected in stacked Spitzer images, so that if they represent a tail of the standard infrared flux density distribution, the distribution is either bimodal, or else has a very long tail.

We have recently (Norris et al. 2007; Middelberg et al. 2008) detected two of them using VLBI, implying that they have AGN cores. The cores are presumably so heavily obscured, or at such a high redshift, that all their dust emission is shifted beyond Spitzer's wavelength coverage. For example, 3C273 at z=7 would have very similar observational characteristics to the IFRS.

Further evidence for an AGN origin comes from the wide range (-2 to +1) of spectral indices. However, these measurements are dogged by amplitude calibration uncertainties of about 10\%, which are typical for synthesis images, and which make it difficult to compare flux densities obtained with different telescopes or different frequencies. It is unacceptable that current radio-astronomical imaging techniques produce 10\% amplitude uncertainties, when signal-to-noise considerations indicate that amplitudes should be accurate to about 1\% or better.

\underline{SKA Lesson 3:} {\em We need to improve the measurement of flux densities from synthesis images if we are to reap the full value of the SKA.}

\section{Conclusion and Future Work}

Because of their ability to penetrate dust and reveal AGNs, deep wide radio surveys are starting to have a profound impact on our understanding of galaxy evolution. 
Such surveys are also lighting the path forward for next-generation radio telescopes such as SKA and its pathfinders. Already, we can learn three lessons from ATLAS. First, dynamic range may be limited by new artefacts, such as those caused by 
rotation of the primary beam. Second, don't extrapolate too far from the local universe to tell you what you'll see at high z (e.g. embedded AGN).
Third, we need to work out how to measure fluxes more accurately from synthesis observations, so that we can intercompare observations, and derive quantities such as spectral index and the radio-to-infrared ratio.

\subsection*{Acknowledgements}

We thank the entire ATLAS team, listed on http://www.atnf.csiro.au/research/deep/team/index.htm, for their contributions to the work summarised here.

\subsection*{References}
Appleton, P. N., et al. 2004, ApJS, 154, 147\\
Beswick, R. J., Muxlow, T. W. B., Thrall, H., and Richards, A. M. S., 2007, astro-ph/0612077\\
Boyle, B., et al., 2007, MNRAS, 376, 1182B\\
Carilli, C., Rawlings, S., 2004, New Astronomy Reviews, Vol.48\\
de Jong T., Klein,U., Wielebinski,R., Wunderlich,E.,1985, AA, 147, 6\\
Deboer, D., et al., 2007, http://www.atnf.csiro.au/projects/askap/\\
Keel, W.C., et al. 2006, AJ, 132, 2233\\
Lonsdale, C. J., et al.\ 2003, PASP, 115, 897\\
Middelberg, E., et al., 2008, accepted by AJ.\\
Norris, R.~P., et al.\  2006, AJ, 132, 2409\\
Norris R. P., et al.\ 2007, MNRAS, 378, 1434\\
Norris R. P., et al.\ 2008, in preparation\\

\end{document}